\documentclass[12p]{article}
\usepackage{latexsym}
\usepackage{color}
\usepackage{amsmath}
\usepackage{epsfig}
\usepackage{hyperref}
 \usepackage{dcolumn}% Align table columns on decimal point
   \usepackage{threeparttable}

\newcommand{\ortala}[1]{\begin{center}#1\end{center}}

\newcommand{\sandd}[1]{\left\langle #1\right\rangle}

\newcommand{\summ}[3]{{{\underset{#1 }{\overset{#2}{\displaystyle\sum}}}#3}}

\newcommand{\re}[1]{(\ref{#1})}

\newcommand{\eq}[2]{\begin{equation}\label{#1}  #2\end{equation}}

\newcommand{\paran}[1]{\left(#1\right)}
\newcommand{\parank}[1]{\left[#1\right]}

\newcommand{\sch}[1]{Schrodinger}

\newcommand{\komb}[2]{\paran{\begin{array}{c} #1 \\ #2 \end{array}}}

\linethickness{0.6pt}

\begin{document}
%\tableofcontents

\ortala{\textbf{Hysteresis Behavior of Anisotropic Heisenberg Model in Thin Film Geometry}}

\ortala{\textbf{\"Umit Ak\i nc\i \footnote{umit.akinci@deu.edu.tr},Musa Atl\i han\footnote{musa.atlihan@outlook.com}} }

\ortala{\textit{Department of Physics, Dokuz Eyl\"ul University,
TR-35160 Izmir, Turkey}}

\section{Abstract}

The effect of the anisotropy in the exchange interaction and film thickness on the hysteresis behavior of the  anisotropic Heisenberg thin film has been investigated with effective field formulation in a two spin cluster using the decoupling approximation. The behaviors of the hysteresis loop area, coercive field and remanent magnetization with the film thickness and anisotropy in the exchange interaction have been obtained.

Keywords: \textbf{Quantum anisotropic Heisenberg model; hysteresis; thin film}

\section{Introduction}\label{introduction}

The magnetic properties of thin films are drastically different from the bulk counterparts. In the presence of the free surfaces  the translational symmetry of the system broken, due to the surface atoms are embedded in an environment of lower symmetry than that of the inner atoms \cite{ref1,ref2}. The most important result of this fact that,  the surface of the thin film could show the ordered phase even if the bulk is in the disordered phase. Depending on the value of the surface exchange interaction, this interesting behavior may occur and this fact has been observed experimentally \cite{ref3,ref4,ref5}. Another interesting magnetic behavior of thin films is dependence of the critical temperature and saturation magnetization value with film thickness.  It was shown that the critical  temperature and the average magnetic moment per atom increases with the increasing thickness of the film \cite{ref6,ref7}.

One class of the films which exhibit a strong uniaxial anisotropy \cite{ref8} can be modeled by Ising model, which is the highly anisotropic limit of the Heisenberg model. This type of model on thin film geometry has been widely studied in literature by means of several theoretical methods. Widely used methods for these systems are  Monte Carlo (MC) simulations \cite{ref9},  mean field approximation (MFA) \cite{ref10} and effective field theory (EFT) \cite{ref11}. In order to mimic the surface effects on the system, different exchange interactions have been defined and the model has been solved within the framework of EFT \cite{ref12}. Also more realistic effects such as amorphisation of the surface  due to environment have been handled within the same framework \cite{ref13}.    There are also higher spin Ising thin films, e.g. spin-1 Ising thin films have been studied \cite{ref14,ref15}.

On the other hand, wide variety of materials do not have strong uniaxial anisotropy. In this case, the model has to include the spin-spin interactions as Heisenberg model. The most simple system with surfaces is semi infinite model, which has only one surface, in contrast to the thin film. Heisenberg model on a semi infinite geometry has been solved using a wide variety of techniques such as Green function method \cite{ref16}, renormalization group technique \cite{ref17}, MFA \cite{ref18}, EFT \cite{ref19,ref20,ref21},
high temperature series expansion \cite{ref22,ref23}. Besides, critical and thermodynamic properties of the bilayer \cite{ref24,ref25} and multilayer \cite{ref26}  systems have been investigated within the cluster variational method in the pair approximation.
Heisenberg model in a thin film geometry has been solved in a limited case. For instance, Green function method \cite{ref27,ref28,ref29}, renormalization group technique \cite{ref30}, EFT \cite{ref31,ref32} and
MC \cite{ref33,ref34}, are among them.

Hysteresis is a common behavior of the most of the physical systems. It
originates from the delay of the response of the system to the driving cyclic
force, and shows itself as a history dependent response. Magnetic hysteresis is one of the most important and interesting
features of the magnetic materials. Hysteresis loop area (HLA), coercive
field (CF) and remanent magnetization (RM) are the parameters that give important clues about the shape of the hysteresis loops.  Also, these hysteresis properties are very important in technological applications such as manufacturing of magnetic
recording media. The RM is defined as residual magnetization which is the
magnetization left behind in the system after an external magnetic field is
removed. CF is defined as the value of the external magnetic field
needed to reverse the sign of the magnetization. On the other hand, HLA is simply the area of the hysteresis loop, which
corresponds to energy loss due to the hysteresis. In our recent work we obtain to the hysteresis behaviors of the anisotropic Heisenberg model on the bulk system within the EFT formulation \cite{ref35}.

The aim of this work is the determine the effect of the anisotropy in the exchange interaction and number of layers of the thin film, on the hysteresis behavior of the  Heisenberg thin film. The phase diagrams and the magnetization behaviors of this model has already been obtained \cite{ref36}.

For this aim, the paper is organized as follows: In Sec. \ref{formulation} we
briefly present the model and  formulation. The results and
discussions are presented in Sec. \ref{results}, and finally Sec.
\ref{conclusion} contains our conclusions.

\section{Model and Formulation}\label{formulation}

\begin{figure}[h]\begin{center}
\epsfig{file=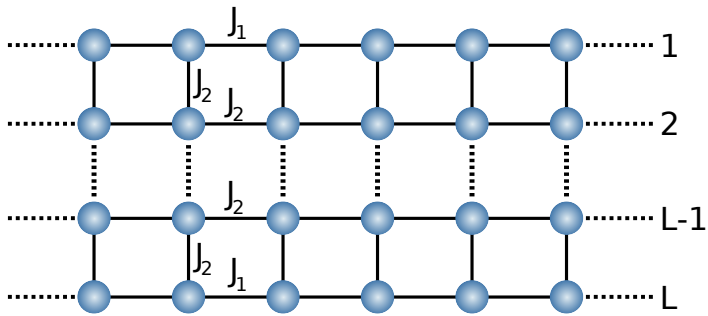, width=10cm}
\end{center}
\caption{Schematic representation of the thin film.}
\label{sek1}\end{figure}

The schematic representation of the thin film can be seen in Fig. \ref{sek1}.  System is infinitely long in $x$ and $y$ directions, while finite in $z$ direction. Thin film can be treated as a layered structure which consist of interacting $L$ parallel layers. Each layer (in $xy$ plane) is defined as a regular lattice with coordination number $z$. When we choose $z=4$, this means that each layer is a square lattice and each nearest neighbor layer have interaction.
The Hamiltonian of the thin film is given by
\eq{denk1}{\mathcal{H}=-\summ{<i,j>}{}{J_{ij}\left[\Delta_{ij}\paran{ s_i^xs_j^x+s_i^ys_j^y}+ s_i^zs_j^z\right]}-\summ{i}{}{H_is_i^z}}
where $s_i^x,s_i^y$ and  $s_i^z$ denote the Pauli spin operators at a site $i$. $J_{ij}$ is the exchange interaction and $\Delta_{ij}$ stands for the anisotropy in the exchange interactions between the nearest
neighbor spins located at sites $i$ and $j$.  Longitudinal magnetic field at a site $i$ is denoted by $H_i$ in Eq. \re{denk1}. The first sum is carried over the nearest neighbors of the lattice, while the second one is over all the lattice sites. The exchange interaction and anisotropy in exchange interaction  between the spins on the sites $i$ and $j$ take the values according to the positions of the nearest neighbor spins. The two surfaces of the film have the intralayer interactions  $(J_{1},\Delta_{1})$. The interlayer coupling between the surface and its adjacent layer (i.e. layers $1,2$ and $L-1,L$) is denoted by $(J_{2},\Delta_{2})$. For the rest of the layers, the interlayer and the intralayer couplings are assumed as $(J_{2},\Delta_{2})$.

In order to get the hysteresis behavior of the anisotropic  Heisenberg thin film,  we use the EFT,   which can provide results that are superior to those obtained within the MFA. This fact comes from the consideration of self spin correlations, which are omitted in the MFA. In an EFT approximation, one constructs a finite cluster. The anisotropic character of the spin-spin interactions cannot be handled in one spin cluster, thus we use two spin cluster approximation here. EFT for the two spin cluster namely EFT-2 formulation \cite{ref37}, was first proposed in Ref. \cite{ref38} for Ising systems. In EFT-2 approximation, we choose
two spins (namely $s_1$ and $s_2$) in each layer and treat interactions exactly in this two spin cluster. In order to avoid some mathematical difficulties,
we replace the perimeter spins of the two spin cluster by Ising spins (axial approximation) \cite{ref39}. After all, by using the differential operator technique
and decoupling approximation (DA) \cite{ref40}, we can get an expression for the magnetization per spin, i.e. $m=\sandd{\frac{1}{2}\paran{s_1^z+s_2^z}}$. In the thin film geometry, the
number of $L$ different representative magnetizations for the system (by following the procedure given in Ref. \cite{ref32}) can be given as,
\eq{denk2}{\begin{array}{lcl}
m_1&=&\sandd{\Theta_{1,1}^3\Theta_{2,2}}f_1\paran{x,y,H_1,H_2}|_{x=0,y=0}\\
%m_2&=&\sandd{\Theta_{2,1}\Theta_{2,2}^3\Theta_{2,3}}f_2\paran{x,y,H_1,H_2}|_{x=0,y=0}\\
m_k&=&\sandd{\Theta_{2,k-1}\Theta_{2,k}^3\Theta_{2,k+1}}f_2\paran{x,y,H_1,H_2}|_{x=0,y=0},k=2,3,\ldots,L-1\\
%m_{L-1}&=&\sandd{\Theta_{2,L-2}\Theta_{2,L-1}^3\Theta_{2,L}}f_2\paran{x,y,H_1,H_2}|_{x=0,y=0}\\
m_L&=&\sandd{\Theta_{2,L-1}\Theta_{1,L}^3}f_1\paran{x,y,H_1,H_2}|_{x=0,y=0}.\\
\end{array}} Here $m_i,(i=1,2,\ldots, L)$ denotes the magnetization of the $i^{th}$ layer. The operators in Eq. \re{denk2} are defined via
\eq{denk3}{
\Theta_{k,l}=\left[A_{kx}+m_lB_{kx}\right]\left[A_{ky}+m_lB_{ky}\right]
} where
\eq{denk4}{\begin{array}{lcl}
A_{km}&=&\cosh{\paran{J_k^z\nabla_m}}\\
B_{km}&=&\sinh{\paran{J_k^z\nabla_m}},\quad k=1,2; m=x,y.
\end{array}
}
The functions in Eq. \re{denk2} are given by
\eq{denk5}{f_n\paran{x,y,H_1,H_2}=\frac{x+y+H_1+H_2}{X_0^{(n)}}\frac{\sinh{\paran{\beta X_0^{(n)}}}}{\cosh{\paran{\beta X_0^{(n)}}}+\exp{\paran{-2\beta J_n^z}}\cosh{\paran{\beta Y_0^{(n)}}}}} where
\eq{denk6}{\begin{array}{lcl}
X_0^{(n)}&=&(x+y+H_1+H_2)\\
Y_0^{(n)}&=&\left[\paran{2\Delta_n J_n^z}^2+(x-y+H_1-H_2)^2\right]^{1/2}\\
\end{array}}with the values $n=1,2$. In Eq. \re{denk5}, we set $\beta=1/(k_B T)$ where $k_B$ is Boltzmann
constant and $T$ is the temperature.

Magnetization expressions given in closed form in Eq. \re{denk2} can be constructed via acting differential operators on related functions.

Differential operator technique depends on writing Eq. \re{denk2} in a polynomial form in $m$ via the exponential differential operator. The effect of the exponential
differential operator to an arbitrary  function $F(x)$ is given by
\eq{denk6}{\exp{\paran{a\nabla}}F\paran{x}=F\paran{x+a}} with any
constant  $a$.

By using Eq. \re{denk3} in Eq. \re{denk2} and converting the hypertrigonometric functions to exponentials, after using the Binomial expansion, Eq. \re{denk2} can be written in the form
\eq{denk8}{\begin{array}{lcl}
m_1&=&\summ{p=0}{3}{}\summ{q=0}{3}{}\summ{r=0}{1}{}\summ{s=0}{1}{}K_1\paran{p,q,r,s}m_1^{p+q} m_2^{r+s}\\
%m_2&=&\summ{p=0}{1}{}\summ{q=0}{1}{}\summ{r=0}{3}{}\summ{s=0}{3}{}K_2\paran{p,q,r,s}m_1^{p+q} m_2^{r+s}\\
m_k&=&\summ{p=0}{1}{}\summ{q=0}{1}{}\summ{r=0}{3}{}\summ{s=0}{3}{}\summ{t=0}{1}{}\summ{v=0}{1}{}K_2\paran{p,q,r,s,t,v}m_{k-1}^{p+q} m_k^{r+s}m_{k+1}^{t+v}\\
%m_{L-1}&=&\summ{p=0}{1}{}\summ{q=0}{1}{}\summ{r=0}{3}{}\summ{s=0}{3}{}K_2\paran{p,q,r,s}m_L^{p+q} m_{L-1}^{r+s}\\
m_L&=&\summ{p=0}{3}{}\summ{q=0}{3}{}\summ{r=0}{1}{}\summ{s=0}{1}{}K_1\paran{p,q,r,s}m_L^{p+q} m_{L-1}^{r+s}\\
\end{array}} where the coefficients are defined by

\eq{denk9}{\begin{array}{lcl}
K_1\paran{p,q,r,s}&=&\komb{3}{p}\komb{3}{q}A_{1x}^{3-p}A_{1y}^{3-q}A_{2x}^{1-r}A_{2y}^{1-s}\times \\
&&B_{1x}^{p}B_{1y}^{q}B_{2x}^{r}B_{2y}^{s}f_1\paran{x,y,H_1,H_2}|_{x=0,y=0}\\
%K_2\paran{p,q,r,s}&=&\komb{3}{r}\komb{3}{s}A_{2x}^{4-(p+r)}A_{2y}^{4-(q+s)}B_{2x}^{p+r}B_{2y}^{q+s}f_2\paran{x,y,H_1,H_2}|_{x=0,y=0}\\
K_2\paran{p,q,r,s,t,v}&=&\komb{3}{r}\komb{3}{s}A_{2x}^{5-(p+r+t)}A_{2y}^{4-(q+s+v)}\times \\
&&B_{2x}^{p+r+t}B_{2y}^{q+s+v}f_2\paran{x,y,H_1,H_2}|_{x=0,y=0}.\\
\end{array}}
These coefficients can be calculated from definitions given in Eq. \re{denk4} by using Eqs. \re{denk5} and \re{denk6}.

Eq. \re{denk8} is a  system of coupled non linear equation system, and it can be solved via usual Newton-Raphson iteration \cite{ref41}. The solution of the system is the longitudinal magnetizations of each layer ($m_i,i=1,2,\ldots,L$). The total longitudinal magnetization ($m$) can be calculated via
\eq{denk10}{m=\frac{1}{L}\summ{i=1}{L}{m_i}.}

We can determine the second order critical point by linearizing the equation system given in Eq. \re{denk8}. Since all the longitudinal magnetizations are close to zero in the vicinity of the second order critical point, the solution of the
\eq{denk11}{\begin{array}{lcl}
m_1&=&\parank{K_1\paran{1,0,0,0}+K_1\paran{0,1,0,0}}m_1+\\
&&\parank{K_1\paran{0,0,1,0}+K_1\paran{0,0,0,1}}m_2\\
m_k&=&\parank{K_3\paran{1,0,0,0,0,0}+K_3\paran{0,1,0,0,0,0}}m_{k-1}+\\
&&\parank{K_3\paran{0,0,1,0,0,0}+K_3\paran{0,0,0,1,0,0}}m_{k}+\\
&&\parank{K_3\paran{0,0,0,0,1,0}+K_3\paran{0,0,0,0,0,1}}m_{k+1}\\
m_L&=&\parank{K_1\paran{1,0,0,0}+K_1\paran{0,1,0,0}}m_L+\\
&&\parank{K_1\paran{0,0,1,0}+K_1\paran{0,0,0,1}}m_{L-1}.\\
\end{array}}
linear equation system will give the second order critical point.

\section{Results and Discussion}\label{results}

The parameters ($J_1^z,\Delta_1$) give the exchange interaction of the nearest neighbor spins at the surfaces, while  ($J_2^z,\Delta_2$) gives the remaining exchange interactions between the nearest neighbor spins of the thin film.

Let us select the  unit of energy as $J$ ($J>0$) and scale the exchange interaction with $J$ as,
\eq{denk12}{
r_i=\frac{J_i^z}{J}, \quad i=1,2
} This defined dimensionless parameters in Eq. \re{denk12} are all positive or zero, since the system has only ferromagnetic interactions.  Let the parameter $q$ to control the ratio of the anisotropy in the exchange interactions between the surfaces and other layers,
\eq{denk13}{
q=\frac{\Delta_1}{\Delta_2}.
}

\subsection{Isotropic Model}

In this case,
\eq{denk14}{
\Delta_1=\Delta_2=1.0.
} In order to see the effect of the ratio of $r_1/r_2$ on the hysteresis behavior of the film, let us choose $r_2=1.0$. The phase diagram of this system has already been obtained \cite{ref36}. The phase diagram of the thin film that have isotropical exchange interaction shows well known interesting behavior. The phase diagrams for different film thickness  ($L$) intersect at a special point which can be denoted by $(r_1^{*},k_BT_c^{*}/J)$ in the ($r_1,k_BT_c/J$) plane.  This fact comes from the surface of the system. Magnetically disordered surface can coexist with a magnetically ordered bulk phase for the values of $r_1$ that provide $r_1<r_1^{*}$ while for the values $r_1>r_1^{*}$, surface can reach the magnetically ordered phase before the bulk. The special point coordinate of the thin film has been obtained as $(r_1^{*},k_BT_c^{*}/J)=(1.345,4.891)$ within the EFT-2 formulation \cite{ref36}, which is nothing but the values that makes the critical temperature of the film independent of the thickness of the film $L$. This critical temperature value is just the
critical temperature of the corresponding bulk system (the system with simple cubic lattice) within the same model \cite{ref37}.

\begin{figure}[h]\begin{center}
\epsfig{file=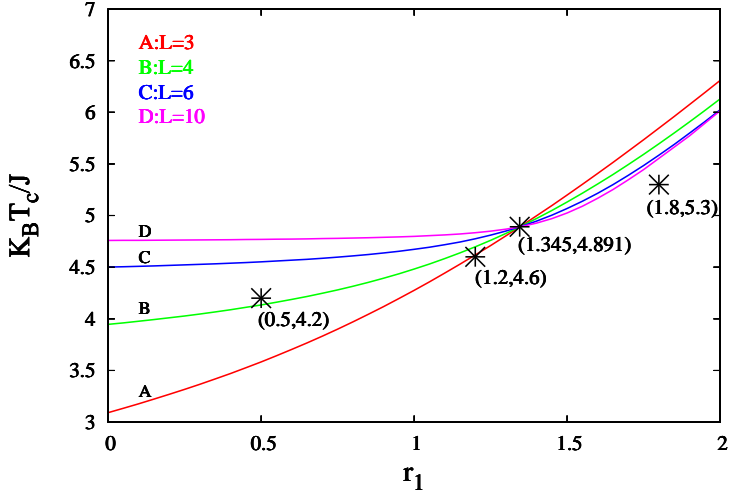, width=10cm}
\end{center}
\caption{The selected values of the temperatures and $r_1$ for the inspection of the hysteresis behaviors in the isotropic model, shown in the phase diagrams of the thin film in isotropic case.
} \label{sek2}\end{figure}

\begin{figure}[h]\begin{center}
\epsfig{file=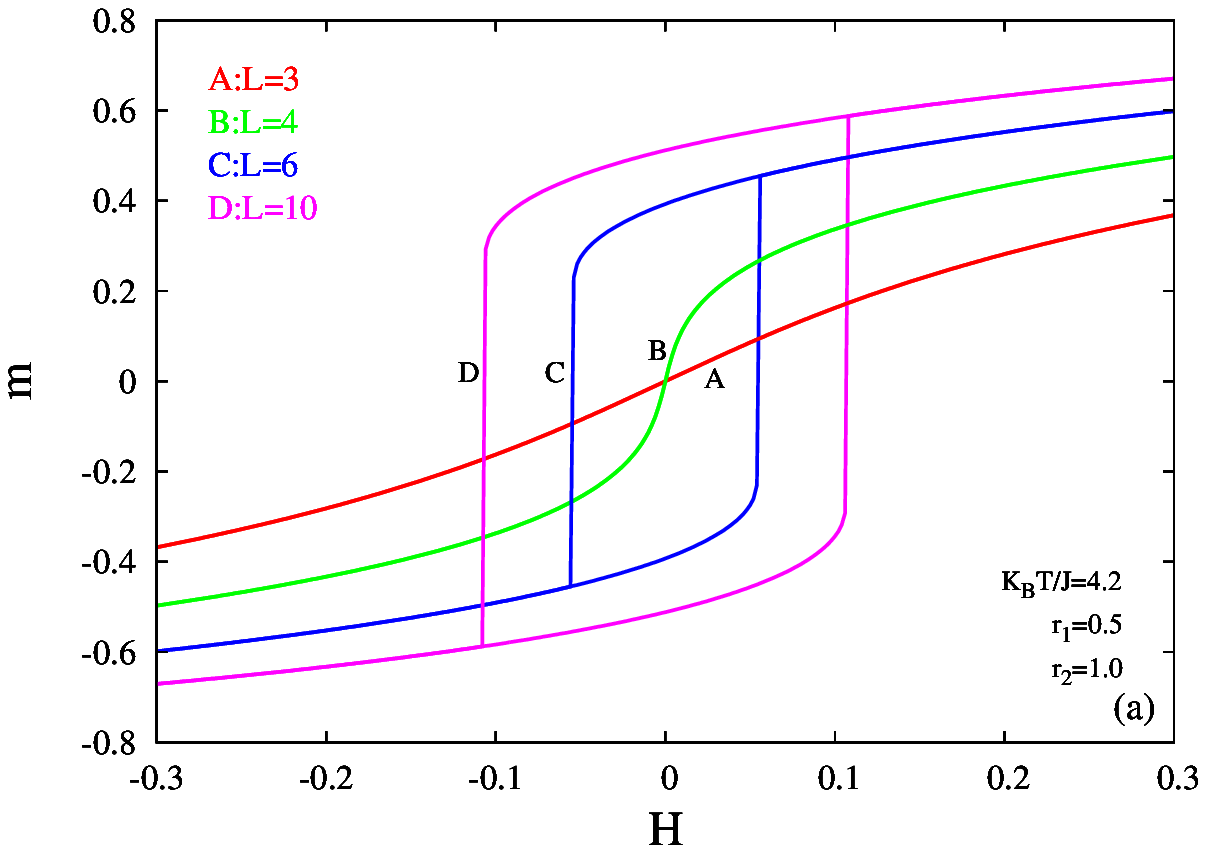, width=6.75cm}
\epsfig{file=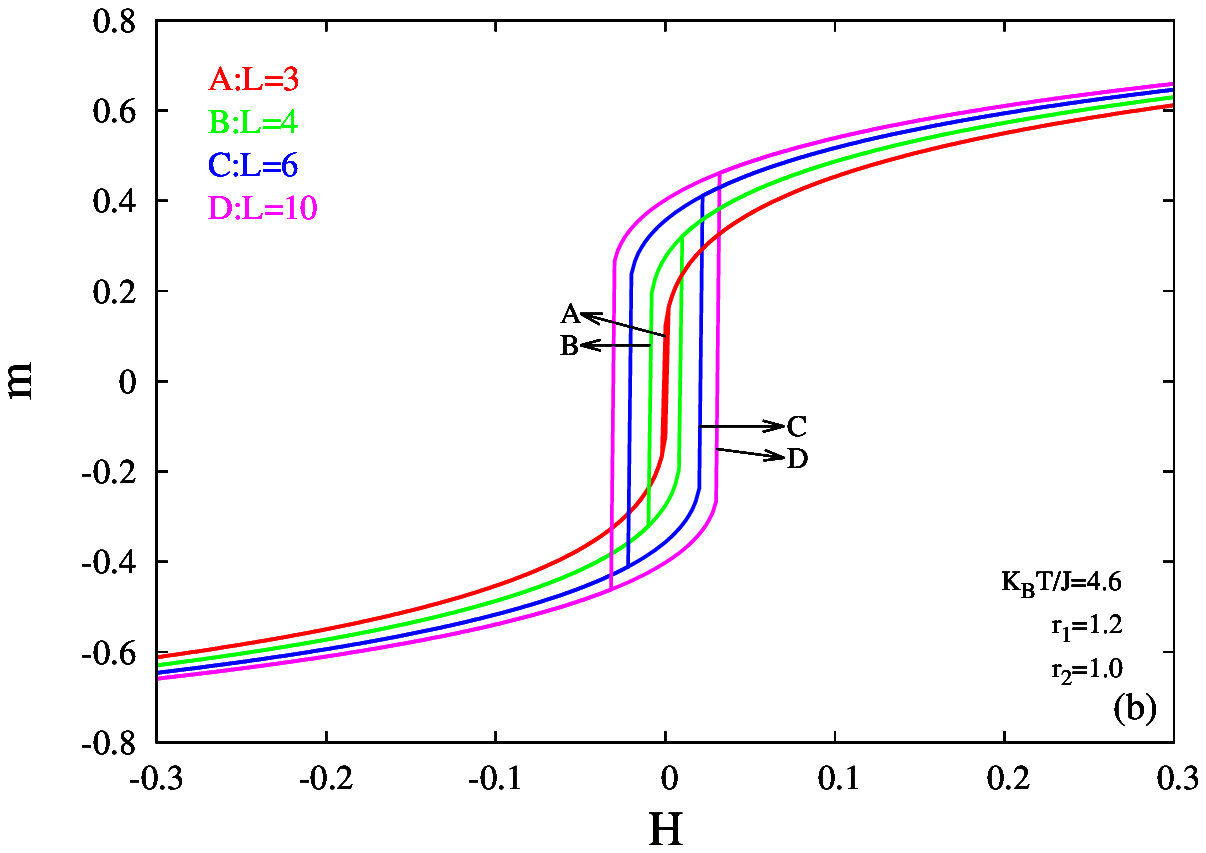, width=6.75cm}
\epsfig{file=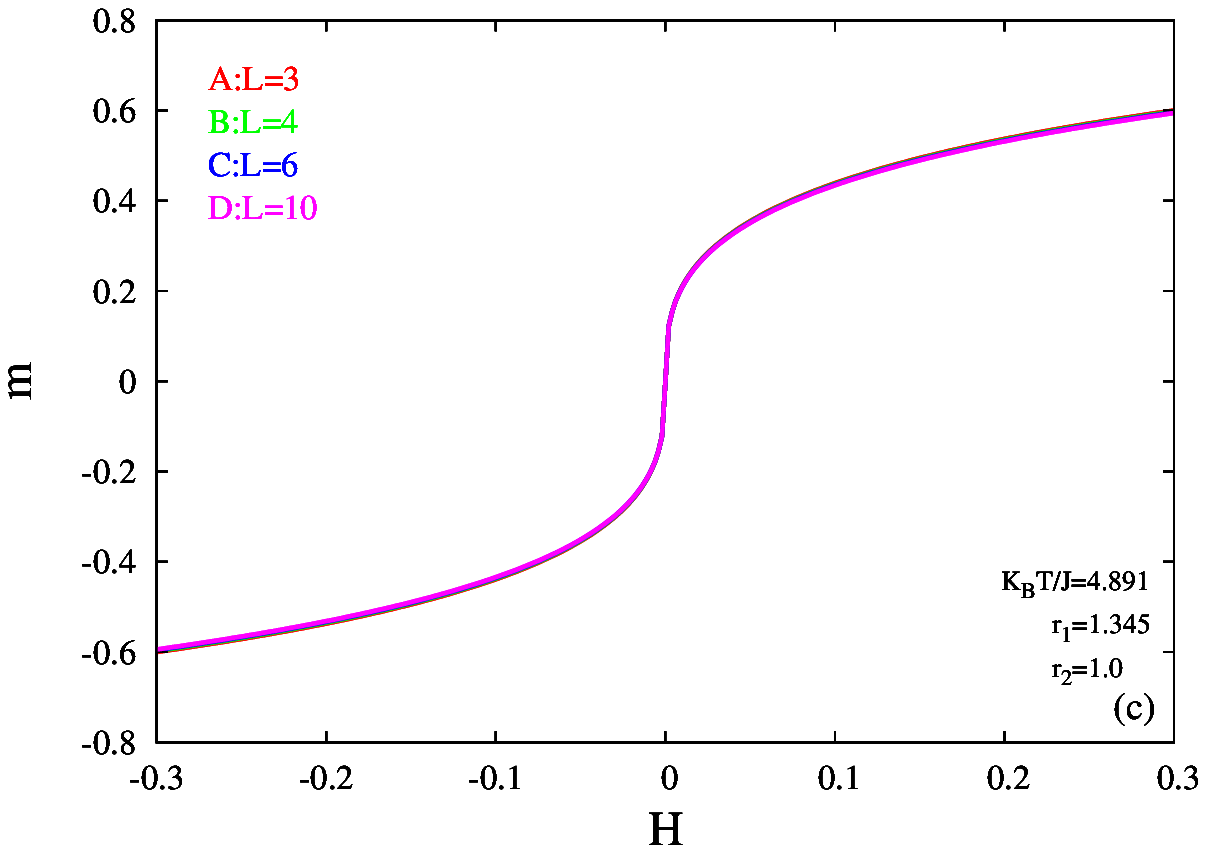, width=6.75cm}
\epsfig{file=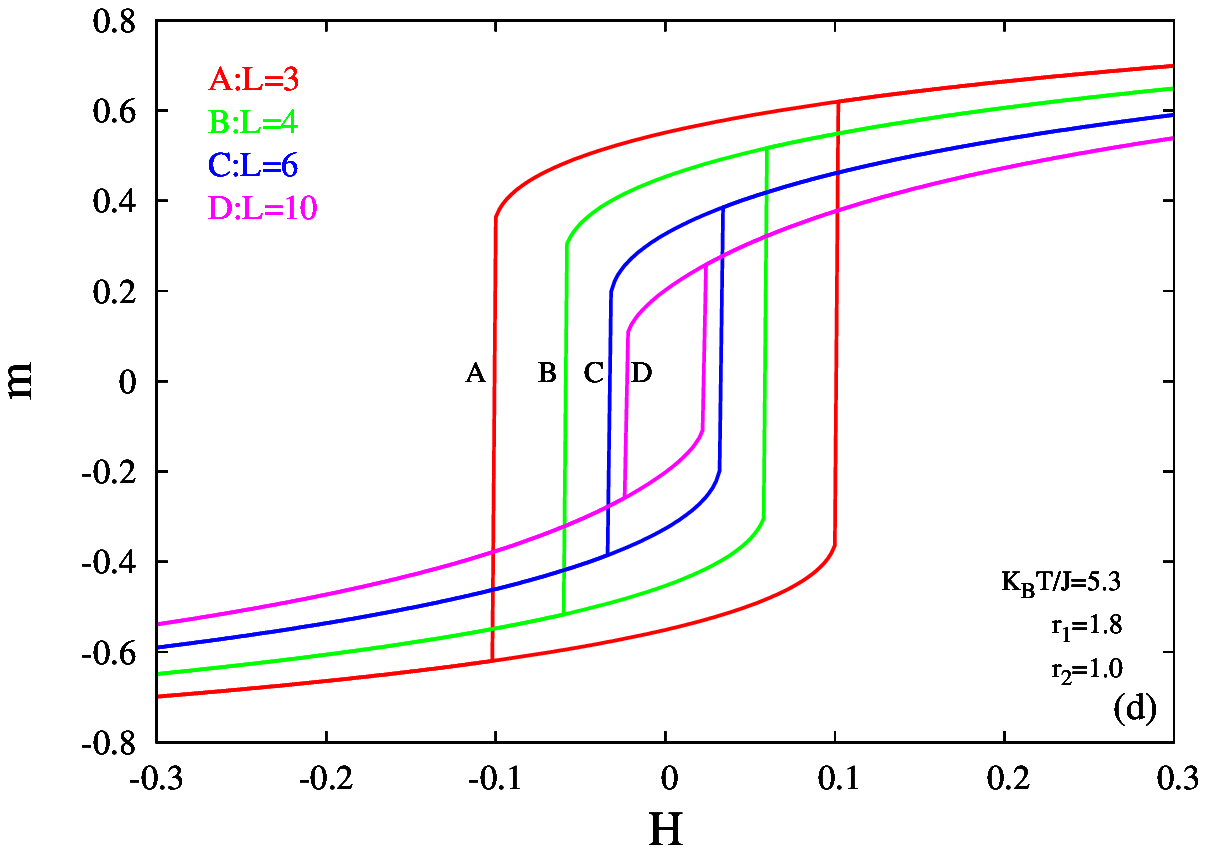, width=6.75cm}
\end{center}
\caption{Hysteresis loops for the isotroic Heisenberg thin film for the $r_2=1.0$ and thicknesses $L=3,4,6,10$, for selected values of the
$(r_1,k_BT/J)=(0.5,4.2), (1.2,4.6), (1.345,4.891)$  and $(1.8,5.3)$.
} \label{sek3}\end{figure}

\begin{figure}[h]\begin{center}
\epsfig{file=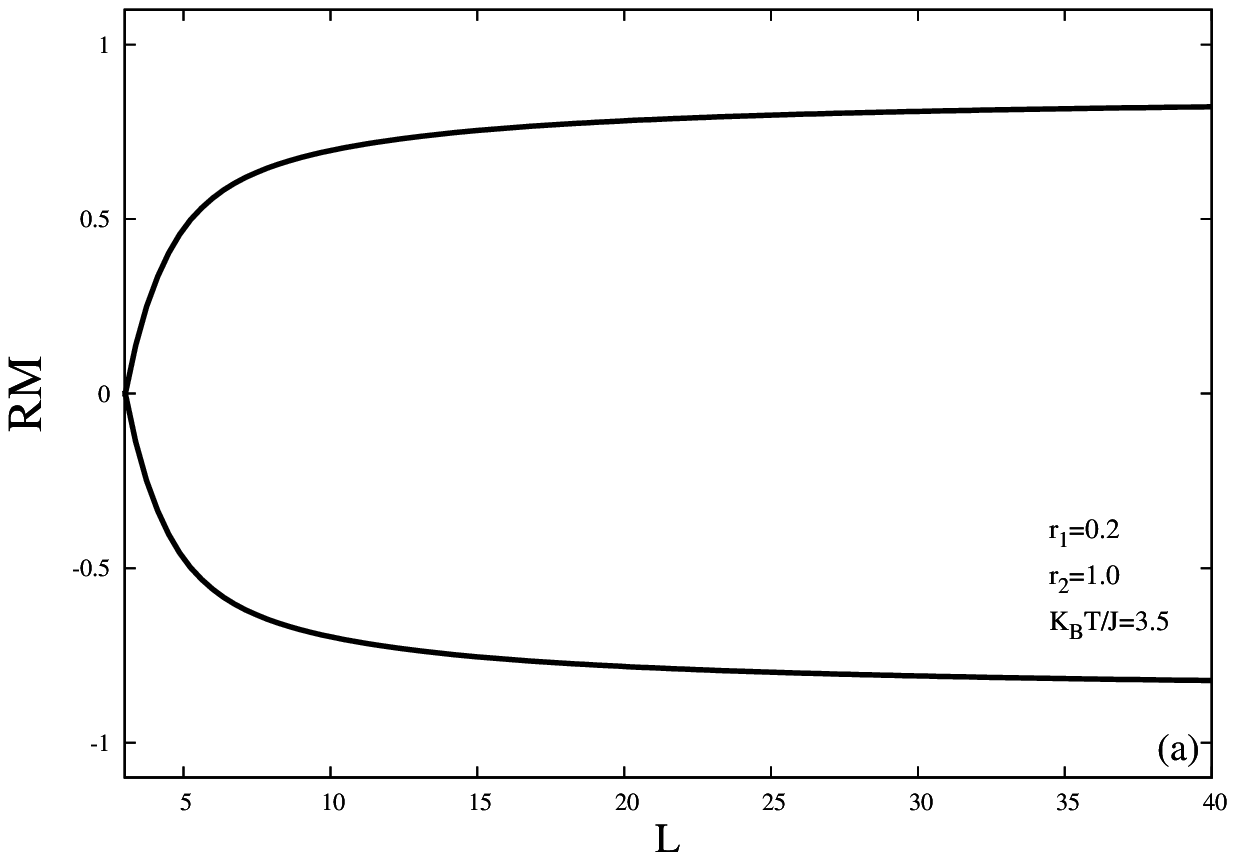, width=6.0cm}
\epsfig{file=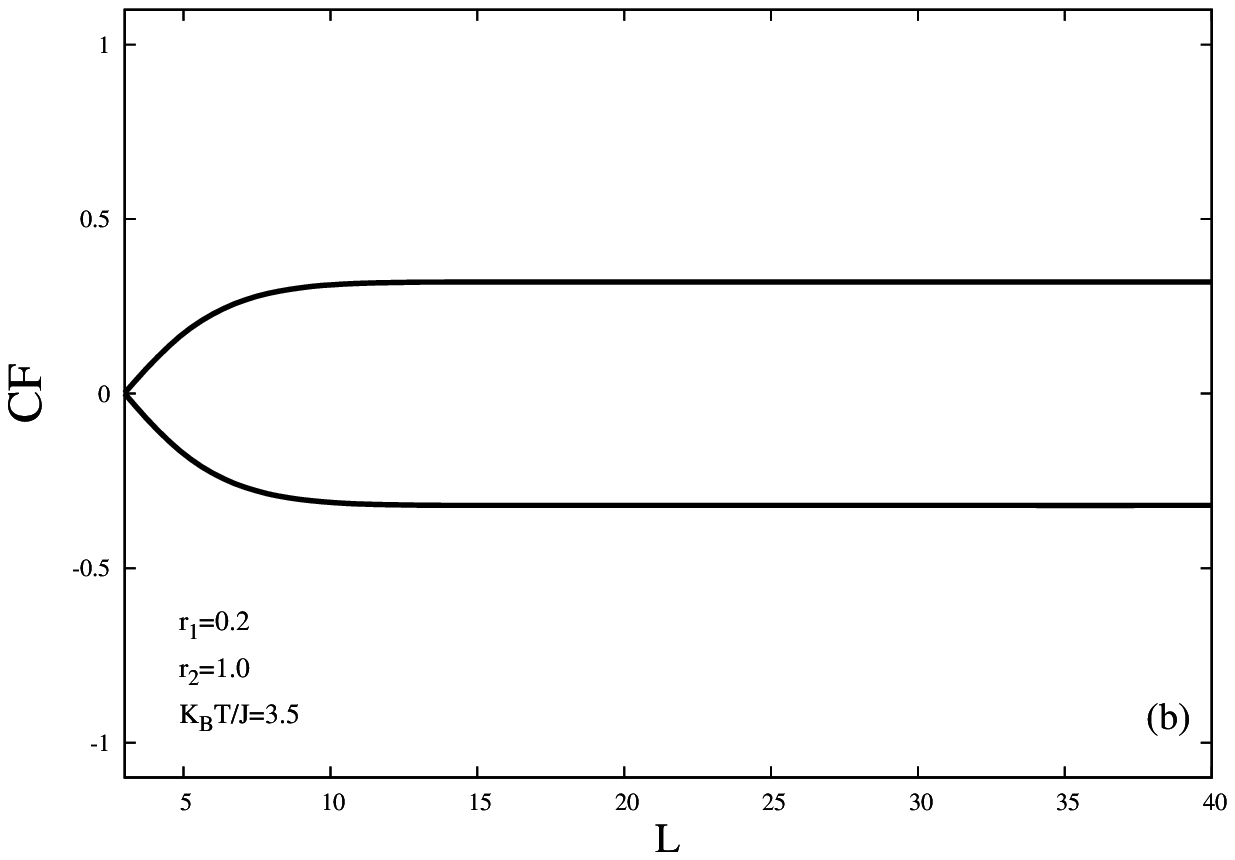, width=6.0cm}
\epsfig{file=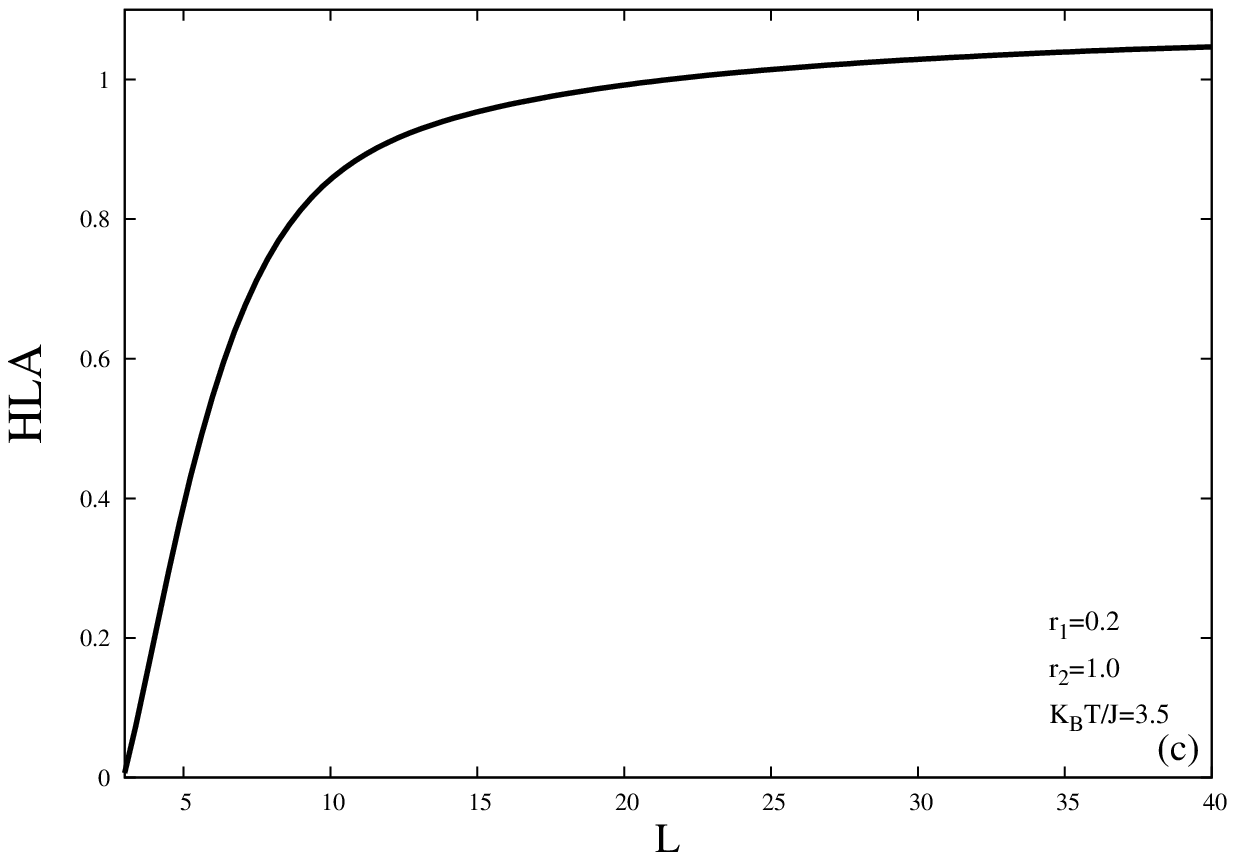, width=6.0cm}
\end{center}
\caption{Variation of the (a) RM, (b) CF and (c) HLA  with the number of layers (film thickness) for selected values of $k_BT/J=3.5$, $r_2=1.0$ and $r_2=0.2$.} \label{sek4}\end{figure}

In order to see the effect of the film thickness $L$ on the hysteresis behavior, we choose four different values of the ($r_1,k_BT/J$), while $r_2=1.0$  as shown in Fig. \ref{sek2}. These behaviors can be seen in Fig. \ref{sek3}, for film thicknesses $L=3,4,6,10$. Typical behavior for the  hysteresis loops with rising thickness for $r_1<r_1^{*}$ is as expected and can be seen in Figs. \ref{sek3} (a) and (b).   When the film gets thicker, ferromagnetic hysteresis loops appear, since rising thickness induce a ferromagnetic phase, as seen in Fig. \ref{sek2}. In a ferromagnetic region, rising thickness give rise to wider hysteresis loops (e.g. compare curves labelled by C and D in Fig. \ref{sek3} (a)). The behavior of the RM, CF and HLA can be seen more clearly in Fig. \ref{sek4} for this region. The parameter values in Fig. \ref{sek4} are $r_1=0.2<r_1^*$, $r_2=1.0$ and  $k_BT/J=3.5$. For this set of values, only the film that has thickness $L=3$ is in the paramagnetic phase, as seen in Fig. \ref{sek2}. Rising thickness, first rises the RM, CF and HLA smoothly. Then, there is no significant change of these values, when the film gets thicker.

At a special point $(r_1^{*},k_BT_c^{*}/J)=(1.345,4.891)$, rising film thickness does not change the hysteresis loops. At this value, all hysteresis loops are the same, as seen in Fig. \ref{sek3} (c). Lastly, when the parameter values selected as $(r_1,k_BT/J)=(1.8,5.3)$, rising film thickness effects on the hysteresis loops reversely in comparison with the parameters that lies on the left of the special point. As seen in Fig. \ref{sek3} (d), hysteresis loops get narrower when the film thickness rises.

All these behaviors can be explained by the special behavior of the thin films namely ordinary and extraordinary phase transition behaviors. For the values that $r_1<r_1^{*}$ bulk part of the film is dominant for the magnetic properties of the film, where ordinary phase transition takes place and thicker films have higher critical temperatures.  On the other hand for the values that provide $r_1>r_1^{*}$, extraordinary transition takes place. In this case surface is dominant, and thicker films have lower critical temperatures and the surface of the film can be ordered before the bulk when the temperature is lowered. Also at any temperature which is below the critical temperature and not so close to the zero, the value of the surface magnetization is reduced in comparison with the bulk for the $r_1<r_1^{*}$, and vice versa. This facts are valid for the isotropic Heisenberg model \cite{ref36} as well as the Ising model \cite{ref12} and this outcome was also proven by more sophisticated
techniques such as series expansion methods and Monte Carlo
simulations \cite{ref42}. We can conclude for this section that, in the  ordinary region (i.e. the parameter values that the system can display ordinary transition, $r_1<r_1^{*}$) rising film thickness enhances the CF, RM and HLA, while the reverse is true for the extraordinary region (i.e. $r_1>r_1^{*}$).

\subsection{Anisotropic Model}

In this case the anisotropies in the exchange interactions are related to each other as given in Eq. \re{denk13} and $\Delta_2=1.0$, i.e.
the interior of the film consists of completely isotropic exchange interactions while the surfaces of the film have the exchange interactions of the Ising type anisotropic ($q=0$), completely isotropic ($q=1$) or XXZ type anisotropic ($0<q<1$), according to the value of $q$.

\begin{figure}[h]\begin{center}
\epsfig{file=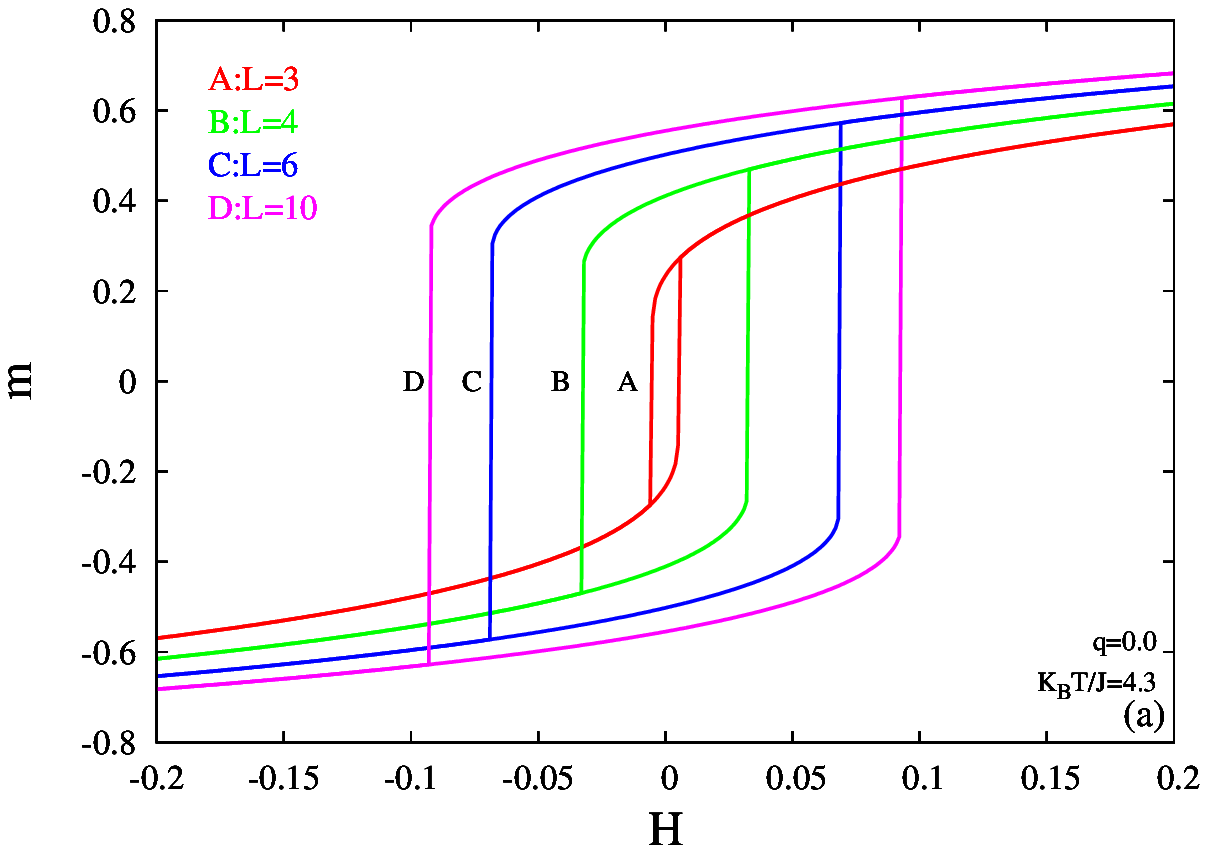, width=6.75cm}
\epsfig{file=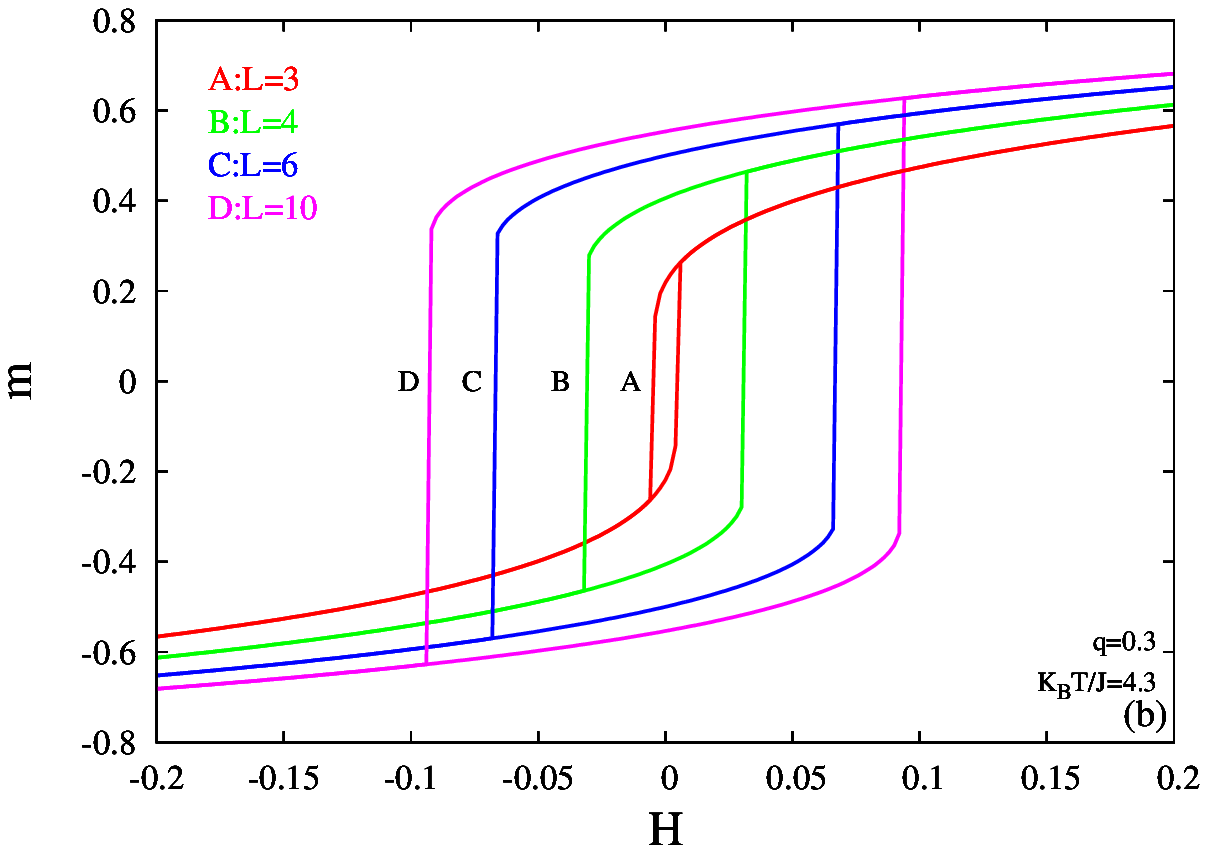, width=6.75cm}
\epsfig{file=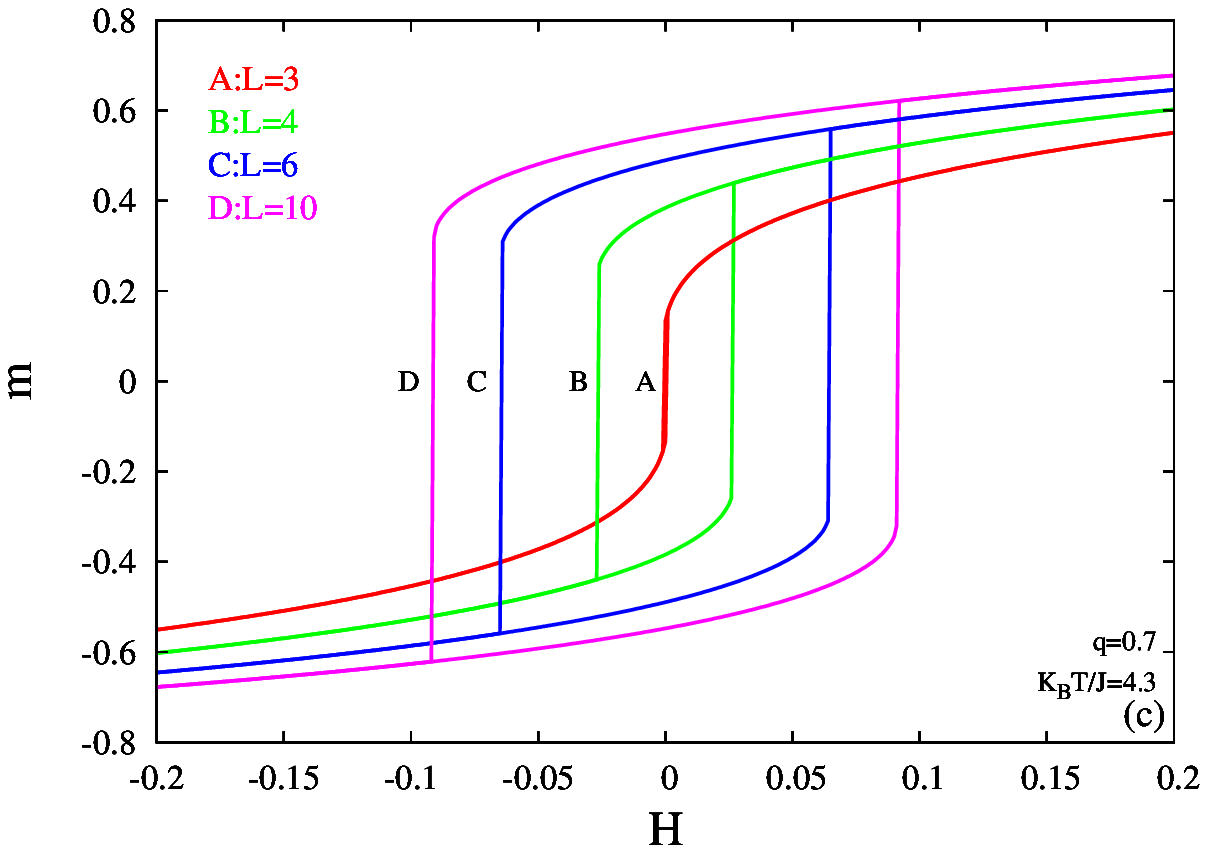, width=6.75cm}
\epsfig{file=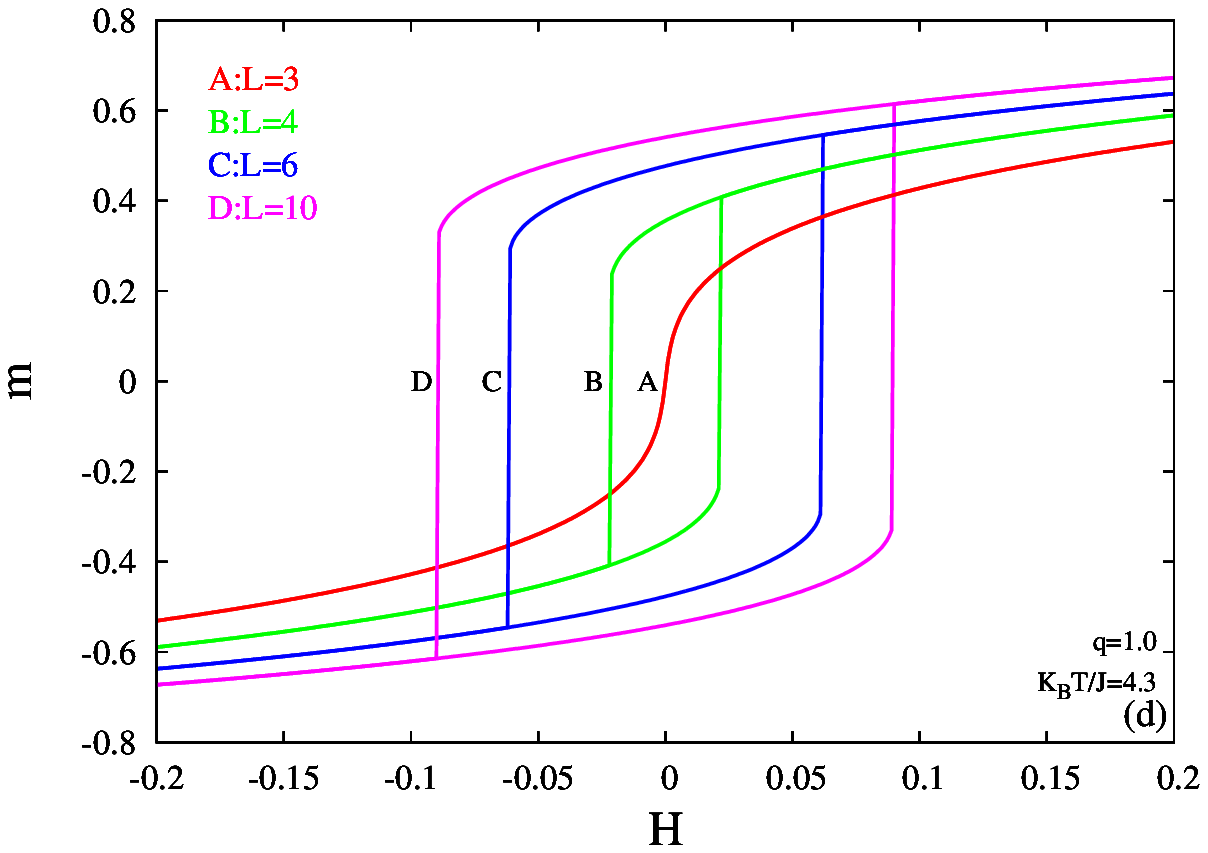, width=6.75cm}
\end{center}
\caption{Hysteresis loops for the anisotroic Heisenberg thin film for the $r_2=1.0$, $k_BT/J=4.3$ and thicknesses $L=3,4,6,10$, for selected values of the  (a) $q=0.0$,   (b) $q=0.3$,   (a) $q=0.7$, and   (a) $q=1.0$.} \label{sek5}\end{figure}

Hysteresis loops of the anisotropic thin film can be seen in Fig. \ref{sek5}. While Fig. \ref{sek5} (a) corresponds to the Ising type surface and Fig. \ref{sek5} (d) isotropic Heisenberg type surface, Figs. \ref{sek5} (b) and (c) are related to the XXZ type anisotropic surface  with the values of $q=0.3$ and $q=0.7$, respectively. The temperature is chosen as $k_BT/J=4.3$ in Fig. \ref{sek5}.

First, all figures exhibit the same behavior with rising film thickness, that is, rising film thickness enlarges the hysteresis loops. This regular behavior is valid for the cases with different types of surfaces (i.e. surfaces that have different anisotropy in the exchange interaction).  The only difference is, the loop labeled by A in Fig. \ref{sek5} (d) which corresponds to the paramagnetic phase for the film thickness $L=3$, while all other loops  are related to the ferromagnetic phase.

On the other hand, when the anisotropy in the exchange interaction of the surfaces of the film is lowered, then the hysteresis loops become narrower (e.g. see the loops labeled by D in Figs. \ref{sek5} (a), (b), (c) and (d), which are the loops for the film thickness $L=6$ for different anisotropy in the exchange interaction of the surfaces of the film). In other words, rising film thickness and rising anisotropy in the exchange interaction of the surfaces (i.e. decreasing $q$), affect the hysteresis loops in the same way, i.e. enlarging the loops.

\begin{figure}[h]\begin{center}
\epsfig{file=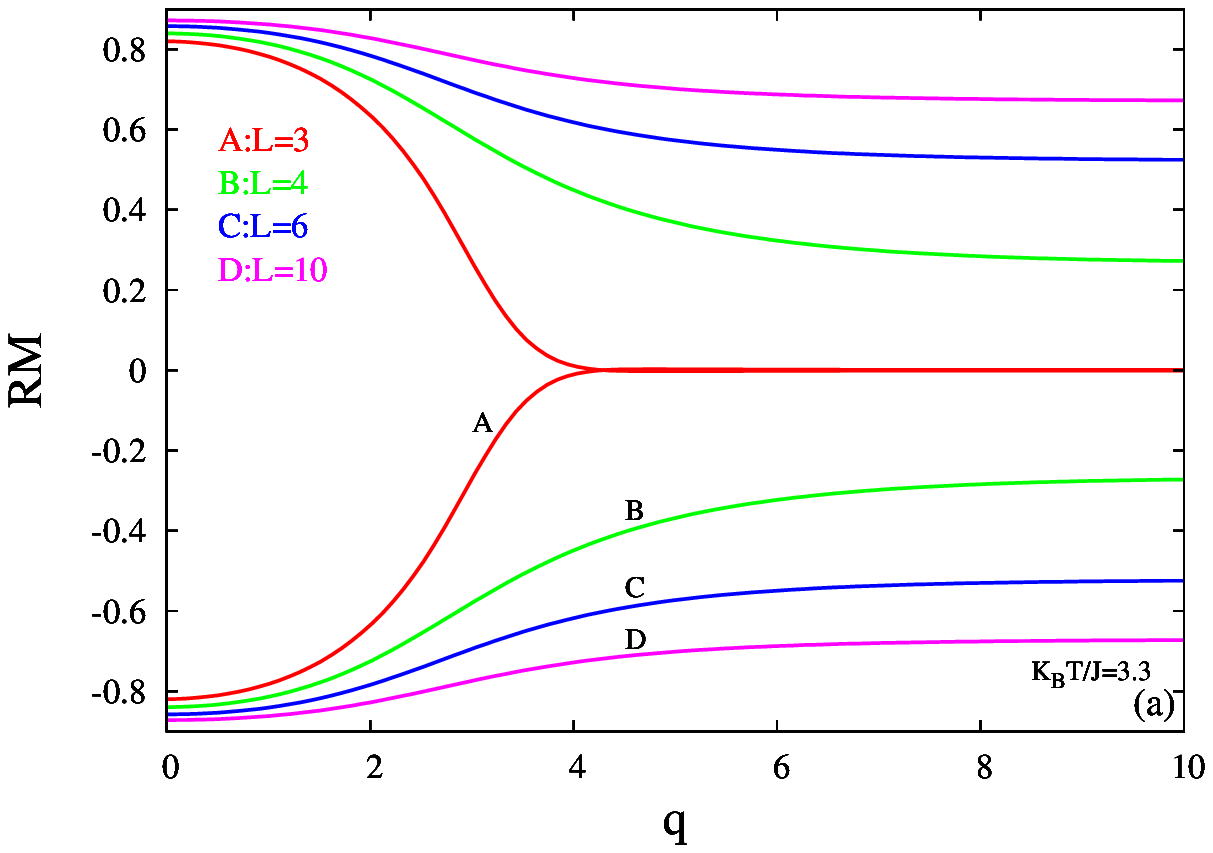, width=6.75cm}
\epsfig{file=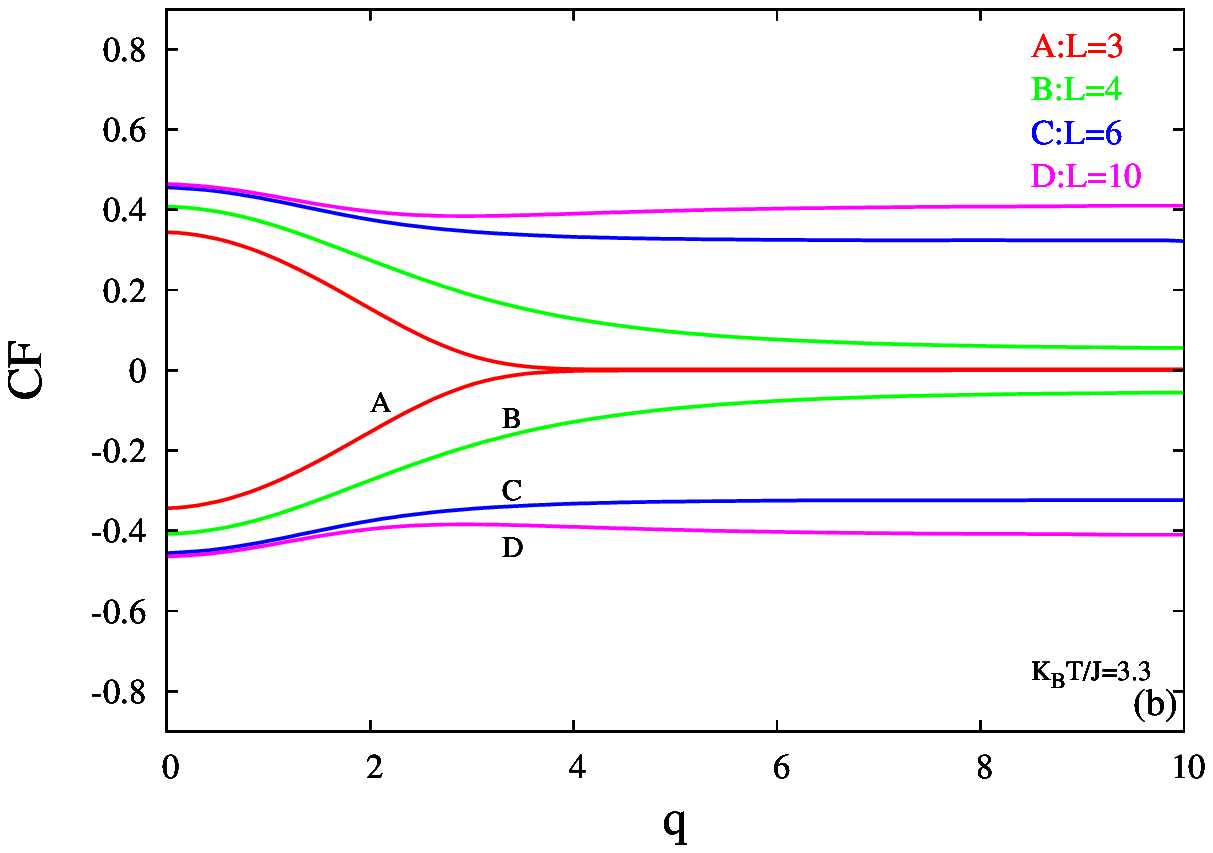, width=6.75cm}
\epsfig{file=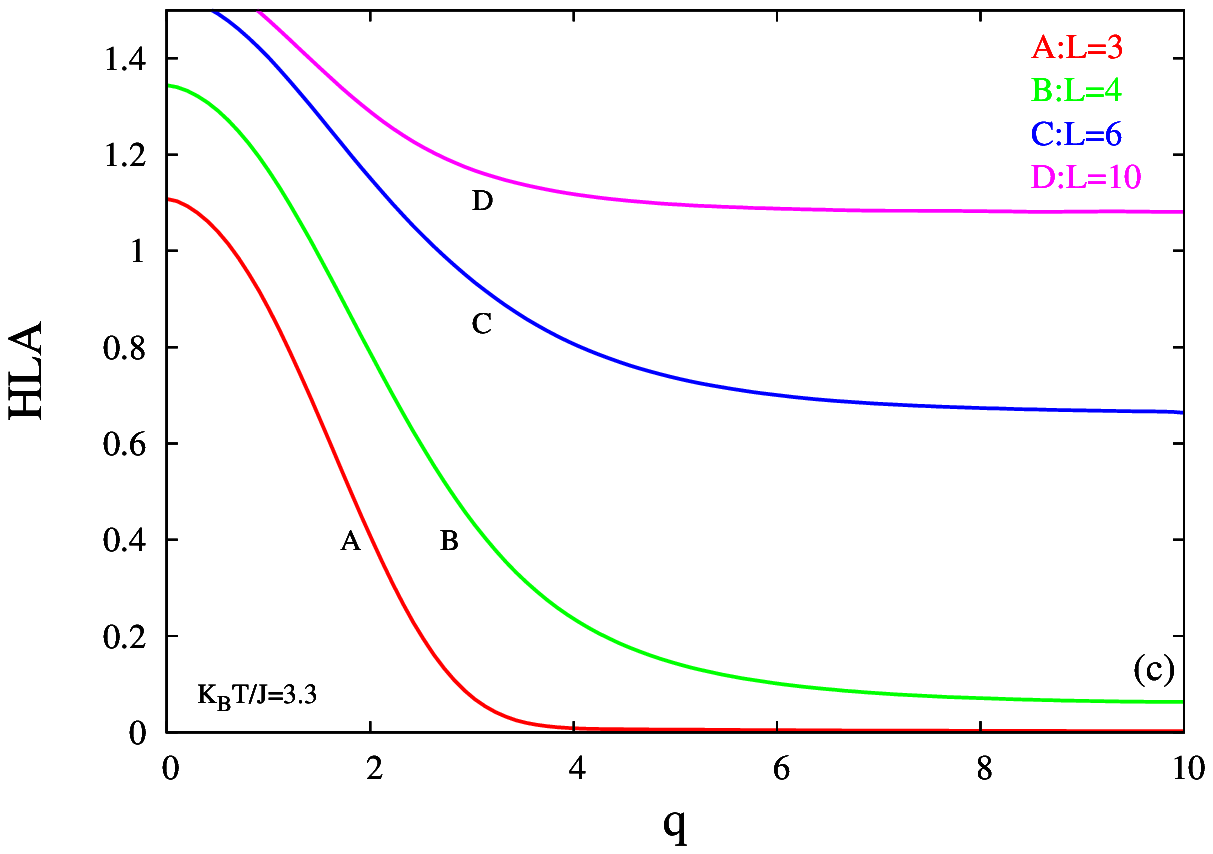, width=6.75cm}
\end{center}
\caption{Variation of the (a) RM, (b) CF and (c) HLA  with the $q$ for the anisotropic model, with number of layers (film thickness) $L=3,4,6,10$ and selected values of $k_BT/J=3.3$, $r_2=1.0$.} \label{sek6}\end{figure}

The shrinking behavior of the hysteresis loops  with decreasing film thickness shows itself in the behavior of the HLA which can be seen in Fig. \ref{sek6} (c), which is the variation of the HLA with $q$ for selected values of  film thickness $L=3,4,6,10$ and temperature $k_BT/J=3.3$.

As shown in Fig. \ref{sek6} (c),  the regular decreasing behavior of the HLA with rising $q$ turns a behavior of staying almost constant. The difference between the HLA of the films that have different thickness, is while the HLA of the film that have thickness $L=3$ getting zero after a specific value 
of $q$ (see curve labeled by A in Fig. \ref{sek6} (c)), the other curves cannot have the value of zero.

The difference between the HLA values of the film that have different thickness is  related to the CF as well as the RM. This fact can be seen in   Figs. \ref{sek6} (a) and (b). In Fig. \ref{sek6} (a) we depict the variation of the RM with $q$ for selected values of $L=3,4,6,10$  and temperature $k_BT/J=3.3$, while the same can be seen in Fig. \ref{sek6} (b) for the CF. It can be seen from these Figs. that the thinner film can have paramagnetic phase when the anisotropy in the exchange interaction of the surface changes.

%% when the surfaces of the film change from the Ising to the isotropic Heisenberg type exchange interaction,
\section{Conclusion}\label{conclusion}

In this work, the effect of the anisotropy in the exchange interaction and the thickness of the film on the hysteresis behavior of thin films has been investigated.
As a formulation, the differential operator technique and DA within the  EFT-2 formulation has been used.

The isotropic model on the thin film displays interesting behavior namely, for the values that provide $r_1<r_1^*$ when the thickness of the film rises, hysteresis loops get wider. This relation gets reversed for the values $r_1>r_1^*$. At a special point (i.e. $r_1=r_1^*$ and $k_BT/J=4.891$) the hysteresis loops are independent of the film thickness.

On the other hand, in the anisotropic case the dependence of the hysteresis loops on the film thickness is more simple than the isotropic case. When the films get thicker, hysteresis loops get wider. This inspection has been made for several different cases by taking the interior of the film as isotropically interacting Heisenberg spins.  The surface of the film that have spins have interaction of the Ising type, anistoropic Heisenberg type (by means of XXZ model) and isotropic Heisenberg type. One conclusion was, when the surfaces of the film have isotropically interacting Heisenberg spins,  hysteresis loops are narrower than the other type of surfaces, when the other parameters (such as the temperature) are kept fixed.   This is because of that, in the former case it is easier to follow the magnetic field than the latter case due to the isotropical spin-spin interaction.

After $q>1.0$ hysteresis loops of the thinner film ($L=3$) disappear due to the transition from the ferromagnetic phase to the paramagnetic one, while the thicker films can stay in an ordered phase.

We hope that the results  obtained in this work may be beneficial form both theoretical and experimental point of view.

%We hope that the results  obtained in this work may be beneficial form both theoretical and experimental point of view.

%% The Appendices part is started with the command \appendix;
%% appendix sections are then done as normal sections
%% \appendix

%% \section{}
%% \label{}

%% References
%%
%% Following citation commands can be used in the body text:
%% Usage of \cite is as follows:
%%   \cite{key}          ==>>  [#]
%%   \cite[chap. 2]{key} ==>>  [#, chap. 2]
%%   \citet{key}         ==>>  Author [#]

%% References with bibTeX database:

\bibliographystyle{model1-num-names}
\bibliography{<your-bib-database>}

\newpage

 \begin{thebibliography}{00}


 \bibitem{ref1}T. Kaneyoshi
Introduction to Surface Magnetism, CRC Press, Boca Raton, Ann Arbor, Boston (1991)



\bibitem{ref2} H. Dosch
Critical Phenomena at Surfaces and Interfaces, Springer, Berlin, Heidelberg, New York (1992)


\bibitem{ref3} C. Ran, C. Jin, M. Roberts,
Journal of Applied Physics  63  (1988) 3667.




\bibitem{ref4} M. Polak, L. Rubinovich, J. Deng,
Physical Review Letters 74    (1995) 4059.



\bibitem{ref5} H. Tang, Physical Review Letters  71  (1985) 444.


\bibitem{ref6}
A.R. Ball, H. Fredrikze, D.M. Lind, R.M. Wolf, P.J.H. Bloemen, M.Th. Rekveldt, P.J. van der Zaag,
Physica B 221  (1996) 388.



\bibitem{ref7}
Yi Li, C. Polaczyk, F. Klose, J. Kapoor, H. Maletta, F. Mezei, D. Riegel,
Physical Review B 53   (1996) 5541.




\bibitem{ref8} K.J. Strandburg, D.W. Hall, C. Liu, S.D. Bader,
Physical Review B  46  (1992) 10818.

%%%%%% enine alansiz  calismalar
%%%% buradan sadece birer guncel calisma secildi




%%% MC enine lan yok s-1/2 cubic ising film
\bibitem{ref9} A. Zaim, Y. El Amraoui, M. Kerouad,  L. Bihc, Ferroelectrics  372  (2008) 3.



%% MFA ising thin film
\bibitem{ref10}F. Aguilera-Granja, J.L. Mor\'an L\'opez,
Solid State Communications 74  (1990) 155.



%% EFT
%% decorated surface, ising thin film, transverse alan yok
\bibitem{ref11}
T. Kaneyoshi, Physica A   293   (2001) 200.

%%% yuzeye farkli omega ve J EFT

%% The effect of a variable surface transverse magnetic field View the MathML source,
%on the order-disorder layering transitions of an Ising spinView the
%MathML source model is investigated using mean field theory (MF) and finite
%cluster approximation (FC) -- ince film


\bibitem{ref12}
T. Kaneyoshi, Physica A 339 (2004) 403.


%%% yuzeye farkli omega ve J EFT bitti




%%%%%%%%% amorphous surface ising thin film



\bibitem{ref13} Youssef El Amraoui, Hamid Arhchoui, S. Sayouri
, Journal of Magnetism and Magnetic Materials 219 (2000) 89.

%%%%%%%%%%%%%%%%%%%%%%%%%% amorphous surface ising thin film bitti






%%% s-1 ince film
\bibitem{ref14}
J.W Jia-Lin Zhong, Chuan-Zhang Yang and Jia-Liang Li, Journal of Physics: Condensed Matter
 3  (1991) 1301.



\bibitem{ref15} L Bahmad,  A Benyoussef, H Ez-Zahraouy, Journal of Magnetism and Magnetic Materials
251  (2002) 115.

%% s-1 ince film bitti !!

%%%%% surface heisenberg basliyor

%%%%%%% semi infinite sistemler

%%% semi infinite, bulk isot, surface XXZ, Heisenberg, GF RPA, fcc, ferro

\bibitem{ref16} S. Seizer, N. Majlis, Physical Review B 27 (1983) 544.


%%%% semi infinite ferro Heisenberg bulk ve surface XXZ, RSRG

\bibitem{ref17} A. M. Mariz, U. M. S. Costa, C. Tsallis, Europhysics Letters
3  (1987) 27.


%%%% semi infinite sc anisotropic quantum heisenberg s-1/2 finite cluster MFA
\bibitem{ref18} J. Cabral Neto, J. Ricardo de Sousa, Physica Status Solidi (b)  212  (1999) 343.




%%  semi infinite sc XXZ heisenberg s-1/2 bimodal rf at surface EFT-2
\bibitem{ref19} Yin-Zhong Wu, Zhen-Ya Li, Solid State Communications  106 (1998) 789.


%%  semi infinite sc XXZ heisenberg s-1/2 EFT-2 ferro
\bibitem{ref20} A. Benyoussef, A. Boubekri, H. Ez-Zahraouy, M. Saber,
Chinese Journal of Physics  37  (1999) 89.


%%%% semi infinite sc anisotropic quantum heisenberg s-1/2  + classical O(n)   EFT-2
\bibitem{ref21} J. Cabral Neto, J. Ricardo de Sousa, Physica A  319  (2003) 319.


%%%% spin-S, ferro, disordered semi infinite, fcc, classical heisenberg, high temperature series expansion

\bibitem{ref22} Huang Zhigao, Feng Qian, Du Youwei, Physics Letters A  372  (2008) 5203.



%%% HTSE, semi infinite, thin film, sc, fcc, Heisenberg

\bibitem{ref23} R. Masrour, M. Hamedoun, A. Benyoussef
, International Journal of Modern Physics B  18  (2010) 3561.



% bilayer and multilayer cluster Variational Method in the Pair Approximation


\bibitem{ref24}  T. Balcerzak , I. Luzniak, Physica A  388  (2009) 357.



\bibitem{ref25} K. Szalowski , T. Balcerzak, Thin Solid Films  534  (2013) 546.


\bibitem{ref26}  K. Szalowski , T. Balcerzak, Physica A  391  (2012) 2197.



%%% ince film


%%% ferro-antiferro thin film  heisenberg anisotropy term of the dipolar interaction type, sc, Green function method

\bibitem{ref27} Diep-The-Hung, J.C.S. Levy, 0. Nagai, Physica Status Solidi (b)  93  (1979) 351.
\bibitem{ref28} Diep-The-Hung, Physica Status Solidi (b)  103  (1981) 809.

%%% acaba bunlar H.T. Diep mi????????????????

%%%%%%%%%%%%%%%%% bcc anti ferro GF, antiferro sistemde beklenemdik davranis bulunmus  thin film

\bibitem{ref29} H.T. Diep, Physical Review B  43 (1991) 8509.

%%% ferro thin film, RG, Heisenberg model with easy-plane single ion anisotropy
\bibitem{ref30} M. Bander, D. L. Mills, Physical Review B  38  (1988) 12015.





%%% ferro thin film klasik n vector ferro thin film, EFT-2

\bibitem{ref31} J. Cabral Neto, J. Ricardo de Sousa
, Journal of Magnetism and Magnetic Materials  268  (2004) 298.

%%% classical and quantum EFT-2
\bibitem{ref32} J. Cabral Neto, J. Ricardo de Sousa, J. A. Plascak
 ,Physical Review B  66  (2002) 064417.


%%% ferro thin film MC

\bibitem{ref33} Huang Zhigao, Feng Qian, Du Youwei
, Journal of Magnetism and Magnetic Materials  269 (2004) 184.

%%% helimagnetic Ho thin film, MC, Heisenberg model with easy-plane single ion anisotropy

\bibitem{ref34} F.Cinti, A.Cuccoli, A.Rettori
, Journal of Magnetism and Magnetic Materials  322  (2010) 1334.


\bibitem{ref35}  \"U. Akýnc\i, Physics Letters A,377, (2013) 1672.


\bibitem{ref36} \"U. Akýnc\i, Thin Solid Films,550, (2014) 602.


%%% EFT-2 on spin-1/2 Heisenberg ferromagnet
\bibitem{ref37} T. Idogaki, N. Ury\^{u},
Physica A 181,  (1992) 173.


%% EFT-2 for ising
\bibitem{ref38} A. Bob\'ak, M. Ja\u{s}\u{c}ur,
Physica Status Solidi (b)  135   (1986) K9.

%% axial appr.
\bibitem{ref39} J. Mielnicki,  G. Wiatrowski, T. Balcerzak,  Journal of Magnetism and Magnetic Materials  71   (1988) 186.



%%%%%%%%%% eft da

\bibitem{ref40}  R. Honmura, T. Kaneyoshi, Journal of Physics C 12 (1979) 3979.



\bibitem{ref41} W.H. Press, S.A. Teukolsky, W.T. Vetterling, B.P. Flannery,
Numerical Recipes: The Art of Scientific Computing (3rd ed.), Cambridge University Press, New York (2007)


\bibitem{ref42} D.P. Landau, K. Binder, Phys. Rev. B 41 (1990) 4633.








%\bibitem{ref81}  T. Kaneyoshi, Acta Phys. Pol. A \textbf{83} (1993) 703.

%% \bibitem must have the following form:
%%   \bibitem{key}...
%%

% \bibitem{}


 \end{thebibliography}

\end{document}